# A new era for dual AGN science with SHARP ☆

P. Severgnini [a,*], C. Vignali [b,c], A. De Rosa [d], E. Portaluri [e], F. Rigamonti [a,f,g], L. Battistini [d,h], L. Bertassi [i,a,g], E. Bertola [j], S. Bianchi [h], E. Bortolas [k], C. Cicone [l], Q. D'Amato [d], R. Della Ceca [a], I. Delvecchio [c], M. Dotti [i,a,g], J. Harms [m,n], I. Lamperti [q,j], F. Mannucci [j], M. Parvatikar [d,o], B. Sala [a,p,g], M. Scialpi [p,q,j], R. Serafinelli [r,s], J. Singh [a], M.V. Zanchetta [j]

[a] *INAF – Osservatorio Astronomico di Brera, via Brera 28, 20121 Milano, Italy*
[b] *Dipartimento di Fisica e Astronomia 'Augusto Righi', Università degli Studi di Bologna, Via Gobetti 93,/2, 40129 Bologna, Italy*
[c] *INAF – Osservatorio di Astrofisica e Scienza dello Spazio di Bologna, Via Gobetti 93,/3, 40129 Bologna, Italy*
[d] *INAF – Istituto di Astrofisica e Planetologia Spaziali, Via Fosso del Cavaliere 100, 00133 Rome, Italy*
[e] *INAF – Osservatorio Astronomico d'Abruzzo, Via Mentore Maggini snc, 64100 Teramo, Italy*
[f] *Como Lake centre for AstroPhysics, via Valleggio 11, 22100, Como, Italy*
[g] *INFN, Sezione di Milano-Bicocca, Piazza della Scienza 3, 20126 Milano, Italy*
[h] *Dipartimento di Matematica e Fisica, Università degli Studi Roma Tre, via della Vasca Navale 84, 00146 Roma, Italy*
[i] *Università degli Studi di Milano-Bicocca, Piazza della Scienza 3, 20126 Milano, Italy*
[j] *INAF – Osservatorio Astrofisico di Arcetri, Largo E. Fermi 5, 50125 Firenze, Italy*
[k] *INAF – Osservatorio Astronomico di Padova, Vicolo dell'Osservatorio, 5, 35122 Padova, Italy*
[l] *Institute of Theoretical Astrophysics, University of Oslo, P.O Box 1029, Blindern, 0315, Oslo, Norway*
[m] *Gran Sasso Science Institute, 67100 L'Aquila, Italy*
[n] *INFN, Laboratori Nazionali del Gran Sasso, 67100 Assergi, Italy*
[o] *Dipartimento di Fisica, Università di Roma Tor Vergata, Via della Ricerca Scientifica, I-00133, Roma, Italy*
[p] *University of Trento, Via Sommarive 14, I-38123 Trento, Italy*
[q] *Università di Firenze, Dipartimento di Fisica e Astronomia, via G. Sansone 1, 50019 Sesto F.no, Firenze, Italy*
[r] *Instituto de Estudios Astrofísicos, Facultad de Ingeniería y Ciencias, Universidad Diego Portales, Avenida Ejército Libertador 441, Santiago, Chile*
[s] *INAF – Osservatorio Astronomico di Roma, Via Frascati 33, 00078, Monte Porzio Catone, Roma, Italy*

## ARTICLE INFO

## ABSTRACT

The search for and the characterization of ultra-compact dual active galactic nuclei (AGN) are among the hottest topics of current extragalactic astrophysics. These systems involve two accreting massive black holes (MBHs) embedded within the same host galaxy, with relative projected separations from a few hundred pc down to a few pc. They are central to understanding hierarchical galaxy formation, black hole growth and demographics, and accretion–feedback coupling in the most extreme interaction phases. Even more compellingly, such tight pairs are the most direct precursors of gravitationally bound binary MBHs (sub-pc scale separation), which are among the loudest emitters of gravitational waves (GWs) in the low-frequency ranges. SHARP will deliver the first statistical census and physical characterization of ultra-compact dual AGN up to cosmic distances, finally bridging the observational gap between kpc-scale pairs and sub-pc GW-emitting binaries, and enabling a breakthrough understanding of MBH growth, feedback and co-evolution across cosmic time.

## 1. Scientific background

The presence of Massive Black Holes (MBHs, $M_{BH} > 10^5 M_\odot$) at the centers of massive galaxies and the existence of correlations between their masses and properties of host galaxy bulges were observationally uncovered by different authors (e.g., Soltan 1982, Kormendy and Richstone 1995, Magorrian et al. 1998, Ferrarese and Merritt 2000,

☆ This article is part of a Special issue entitled: 'SHARP Science Book' published in New Astronomy.

* Corresponding author.
*E-mail address:* paola.severgnini@inaf.it (P. Severgnini).

Gebhardt et al. 2000, Ferrarese and Ford 2005, Bandara et al. 2009, Kormendy and Ho 2013, Savorgnan et al. 2016).

Following the $\Lambda$ Cold Dark Matter cosmological predictions (White and Frenk, 1991; Navarro et al., 1996; Cole et al., 2000), galaxies grow hierarchically through mergers and the MBHs hosted in their nuclei sink to the center of the merger remnant mainly due to dynamical friction against stars, gas, and dark matter. According to simulations, interactions involving two or more galaxies efficiently disturb the gas reservoir, causing it to lose stability and angular momentum, thus flowing inward toward the MBH, where it can trigger Active Galactic Nucleus (AGN) activity (Di Matteo et al., 2005; Hopkins et al., 2006, 2008). In particular, the triggering of AGN is expected to peak in advanced stages of major-merger events, when the stellar bulge separations are <2 kpc (Blecha et al., 2013). Thus, according to theoretical predictions, if MBHs are ubiquitously present at the center of galaxies, kpc and sub-kpc dual AGN (systems in which both MBHs are active and embedded within the same remnant host galaxy, see De Rosa et al., 2019, for a review) are expected to occur in a large fraction of merging galaxies in the final stage of their approach (e.g., Colpi and Dotti, 2011; Koss et al., 2012; Weigel et al., 2018, but see also Cisternas et al., 2011).

During dual AGN formation and evolution, both positive and negative feedback processes can occur, with potentially significant impact on the interstellar and circum-galactic medium (e.g. Koss et al., 2012; Blecha et al., 2013; Mezcua et al., 2014). In principle, the estimate of the true occurrence rate of dual AGN and their physical properties (BH mass, luminosity, accretion rate, etc.) based on observations would provide a powerful constraint on galaxy evolution models and directly probe the final phases of the merger process (Van Wassenhove et al., 2012; Blecha et al., 2013; Colpi, 2014). Crucially, dual AGN systems are the most direct precursors of gravitationally bound binary MBHs (sub-pc separations), which are expected to be among the strongest emitters of low-frequency gravitational waves (GWs), detectable with pulsar timing arrays (PTAs; EPTA Collaboration et al., 2023) and with future space-based and moon-based observatories, such as Laser Interferometer Space Antenna (LISA, see Amaro-Seoane et al., 2023) and Lunar Gravitational Wave Antenna (LGWA, see Harms et al., 2021; Ajith et al., 2025), respectively.

In the next sections (Sections 2 and 3), we summarize the current observational status of dual AGN studies and highlight the critical separation scales that remain inaccessible to present facilities. Particular emphasis is placed on ultra-compact dual AGN, here defined as systems hosting two accreting MBHs within a few hundred pc down to a few pc relative separation, embedded in a single host galaxy. These tight systems remain extremely difficult to detect and impossible to spectroscopically characterize using present instrumentation across cosmic time. Yet, as already quoted above, they are the populations required to address several key open questions, from understanding whether mergers drive MBH growth, to quantifying feedback in the latest coalescence phases, and to constraining the formation pathway of GW-emitting MBH binaries. Accessing this regime is therefore essential to move from theoretical expectations to an empirical framework.

In Sections 4 and 5, we discuss how the advent of SHARP[1] (Saracco et al., 2024; Mahmoodzadeh et al., 2025), the concept study of a multi-mode near-IR spectrograph designed for MORFEO (Ciliegi et al., 2024) at the Extremely Large Telescope[2] (ELT), will open the discovery space of ultra-compact dual AGN, enabling a systematic exploration of this crucial phase of MBH evolution and providing critical electromagnetic context for future low-frequency GW detections.

Throughout the paper, we adopt a $\Lambda$CDM cosmology with $H_0$ = 69.6 km s$^{-1}$ Mpc$^{-1}$, $\Omega_m = 0.3$ and $\Omega_\Lambda = 0.7$.

[1] https://sharp.brera.inaf.it/

[2] https://elt.eso.org/

## 2. From galaxy merger to kpc dual AGN

In broad terms, the different stages of gravitational interaction are: galaxy pairs (tens of kpc separation), dual MBHs (from kpc to few pc separation), gravitational bound binary MBHs (pc/sub-pc separation), and the final collapse. Many works have studied galaxy mergers using multi-band imaging and spectroscopic diagnostics (e.g. Sanders et al., 1988; Ellison et al., 2008, 2011; Scudder et al., 2012; Jin et al., 2021), but very few sample the compact end of the merger sequence, where both MBHs are active. Numerous systems have now been detected with projected separations of 20–60 kpc or more (e.g. Koss et al., 2012; Liu et al., 2011; De Rosa et al., 2018, 2019; Hou et al., 2020; De Rosa et al., 2023; Barrows et al., 2023; Perna et al., 2023, 2025; Battistini et al., 2026), and a growing number (about 50) at z<0.1 have been confirmed with separations below 10 kpc (e.g. Liu et al., 2010; Shen et al., 2011; Liu et al., 2013; Comerford et al., 2015; Fu et al., 2015; Satyapal et al., 2017; Hou et al., 2019; Pfeifle et al., 2019a,b; Chen et al., 2022; De Rosa et al., 2023, and references therein).

Nevertheless, the late stages of the merger sequence remain comparatively poorly explored. At projected separations below a few kpcs, resolving the two AGN becomes increasingly challenging because sub-arcsec angular resolution is required, particularly at intermediate and high redshift. In recent years, high-resolution programs with the Hubble Space Telescope - HST (Xu and Komossa, 2009; Fu et al., 2012; Wang et al., 2026) and large surveys using Gaia (Lemon et al. 2018, 2019, 2023; varstrometry technique: Shen et al. 2019, Hwang et al. 2020, Chen et al. 2023, Wang et al. 2023, Uppal et al. 2024, Chen et al. 2025; Gaia Multi-Peak, GMP, technique: Mannucci et al. 2022, Ciurlo et al. 2023, Mannucci et al. 2023, Scialpi et al. 2024, 2026; see also D'Amato et al. 2026 for astrophysically relevant "contaminants" in this quest) have significantly increased the sample of candidate (and, in several cases, also confirmed) dual AGN down to kpc scales, up to z∼3. In this regard, also the James Webb Space Telescope - JWST, despite being characterized by a small field of view, has been able to provide some dual AGN candidates and bona-fide dual AGN with a separation of a few kpc up to ≈30 kpc at Cosmic Noon (Perna et al., 2023, 2025). The recent wide-field survey carried out by Euclid (Cuillandre et al., 2025; Fabricius et al., 2025; Ulivi et al., 2025) is further expanding this effort by identifying galaxies hosting multiple nuclear components at similar separations. These include candidate secondary nuclei which may host MBHs but do not necessarily correspond to dual AGN. Although the number of confirmed dual AGN at the kpc-scale separation is still relatively small (see, e.g., Wang et al. 2026, Pfeifle et al. 2025 for an updated compilation), it is expected to increase rapidly over the next few years thanks to the combined efforts of these surveys. Current Integral Field Unit (IFU) facilities provide the crucial angular resolution and sensitivity required to study dual AGN candidates with relative separations of a few kpc up to high redshift (z>2). These facilities include the Multi Unit Spectroscopic Explorer (MUSE, Bacon et al., 2010) and the Enhanced Resolution Imager and Spectrograph (ERIS, Davies et al., 2023) at the Very Large Telescope (VLT), the OH-Suppressing Infra-Red Imaging Spectrograph (OSIRIS, Larkin et al., 2006) at Keck Observatory, and the Near Infrared Spectrograph (NIRSpec, Jakobsen et al., 2022) onboard the JWST. In this regime, IFU observations allow us to confirm or rule out their dual nature, estimate their occurrence rate, feeding/feedback processes, and derive fundamental physical properties of each BH, including luminosity, mass, accretion rate, and obscuration. These growing samples will finally enable for the first statistically significant comparison between observed dual-AGN rate and model predictions for systems at kpc separations.

What remains almost completely unconstrained is the ultra-compact sub-kpc dual AGN phase, the transitional stage between the kpc-scale duals accessible today and the final pre-coalescence binary MBH phase.

## 3. Ultra-compact dual AGN

### 3.1. Why they matter

The ultra-compact sub-kpc dual AGN are expected to be the most intense stages of MBH growth, dynamical interaction, inflow, and feedback. Indeed, the physical processes occurring in the ultra-compact phases can contribute up to ∼60% of the total MBH mass assembly (Treister et al., 2012; Schawinski et al., 2015), suggesting that these systems trace the most efficient accretion episodes in the life cycle of a BH. Studying AGN pairs at such tight separations therefore provides unique insight into how galaxy interactions trigger nuclear activity, drive gas inflows, and shape the physical conditions of the host during the final stages of merging (Scudder et al., 2012). Since in this regime the triggering of nuclear activity is expected to be at its highest, powerful winds/outflows can be produced as the result of joint feedback effects from star formation, AGN activity, and merger (see e.g., Yoshida et al., 2016; Cicone et al., 2018; Kollatschny et al., 2020; Carlsen et al., 2026).

As quoted in Section 1, the forthcoming detection of low-frequency GWs, spanning the deci-Hz to nano-Hz regime with LGWA, LISA and PTAs, heightens the urgency to constrain the formation pathways and timescales of MBH binaries. Current estimates of GW sources rely on parameterized merger-rate models and empirical scaling relations (e.g., Buchner et al., 2019), resulting in large systematic uncertainties that may span orders of magnitude (Bonetti et al., 2018, 2019). Ultra-compact dual AGN provide the only direct window into MBH pairs in the final stages of galaxy merging before binary formation, and thus an essential laboratory for predicting the rate and characteristics of future GW events. In this framework, SHARP will provide the statistical census and spectroscopic characterization needed to bridge the gap between currently accessible kpc-scale dual AGN and the sub-pc binary regime. Facilities such as GRAVITY+ (see, e.g., Gravity Collaboration et al. 2017, Gravity+ Collaboration et al. 2026) will instead enable detailed studies of a smaller number of bright systems deeper into the gravitationally-bound binary phase, where the dynamics of the BLR and, possibly, deviations from a purely virial behavior will be investigated.

### 3.2. Candidate selection

In spite of their relevance, directly accessing the ultra-compact regime is particularly challenging; the projected distance between the two accreting nuclei rapidly approaches or falls below the angular resolution limit of most of the current facilities, especially beyond the local Universe.

Very-long-baseline interferometry (VLBI) and complementary observations with Atacama Large Millimeter/sub-millimeter Array (ALMA) can achieve resolution of a fraction of parsec at radio and (sub-)millimeter wavelengths, respectively. However, neither technique alone provides a complete and unambiguous characterization of the detected components, which still requires spatially resolved optical/NIR spectroscopic information. VLBI faces major practical limitations: the technique often resolves out diffuse radio emission (Xu et al., 2024), thus effectively biasing the dual AGN census against systems hosting extended radio structures such as jets or outflows. Furthermore, so far VLBI has been successfully applied only to relatively bright sources (≥1 mJy @ 1.4 GHz), both for local and distant dual AGN at different separations (e.g., Rodriguez et al., 2006; Deane et al., 2014; Spingola et al., 2019; Glikman et al., 2023; Schwartzman et al., 2025). Moreover, two VLBI components alone might mimic a dual AGN system while being a core+jet of a single source. Similarly, while ALMA provides crucial information on the cold gas distribution and kinematics, it cannot by itself robustly distinguish dual accreting nuclei from complex gas structures associated with a single AGN. Hence, a spectroscopic characterization is essential.

To date, one of the strategies used to search for dual AGN at sub-kpc separation has relied on the spectroscopic selection of double-peaked narrow optical emission lines, interpreted as potential signatures of two distinct Narrow Line Regions (NLRs) associated with separate MBHs (Wang et al., 2009; Smith et al., 2010; Liu et al., 2010; Ge et al., 2012; Fu et al., 2012; McGurk et al., 2015; Nevin et al., 2016). Although double-peaked line profiles are degenerate (similar features can arise from single-AGN gas kinematics, such as rotating disks, biconical outflows, jet–cloud interactions, and disturbed ionized gas, Heckman et al. 1984, Whittle et al. 2005, Rosario et al. 2010, Gabányi et al. 2017, Xu and Komossa 2009), this method has produced valuable samples of candidates. It remains one of the few ways to flag sub-kpc systems in large surveys. NLRs typically extend over a few hundred parsecs; when the separation between the two MBHs falls below this scale, their NLRs first overlap and then merge into a single ionized region. As a consequence, double-peaked searches cannot efficiently probe the ultra-compact regime when separations fall within or below the characteristic NLR size.

When the AGN are not obscured, it is possible to observe broad emission lines (BELs, full width at half maximum, FWHM, > 1500 km/s) originating in the broad-line regions (BLRs) at ∼0.01 pc from the central MBH. In principle, in ultra-compact dual AGN, each black hole may retain its own BLR with limited tidal distortion, potentially giving rise to double-peaked BELs shifted in velocity with respect to the narrow emission lines (NELs) (Tsalmantza et al., 2011; Eracleous et al., 2012; Ju et al., 2013; Shen et al., 2013; Wang et al., 2017). However, before the formation of a gravitationally bound MBH binary, the expected relative velocity offset between the two BLRs is likely to be only a few hundred km $s^{-1}$, significantly smaller than the typical BEL widths. As a result, detecting such velocity shifts is observationally extremely challenging, even with high-quality spectra, and requires assumptions on line decomposition that are difficult to robustly justify. Moreover, asymmetric and/or double-peaked BEL profiles are not unique signatures of dual AGN. Similar features can arise from single-AGN BLR kinematics, including eccentric BLRs (Eracleous et al., 1997), deviations from axis-symmetric emissivity distributions such as hot spots (Jovanović et al., 2010) and spiral patterns (Storchi-Bergmann et al., 2017; Rigamonti et al., 2025; Ward et al., 2025; Sottocorno et al., 2026). In addition, even intrinsically symmetric BLRs may appear asymmetric due to partial dust obscuration (Gaskell and Harrington, 2018) and in the case of BLR associated with a recoiling MBH (Volonteri and Madau, 2008).

A complementary time-domain approach for the selection of ultra-compact dual AGN has recently emerged. It is well known that AGN light curves are intrinsically stochastic, displaying correlated ("red") noise variability over a wide range of timescales (see Paolillo and Papadakis, 2025, for a review). In the optical/UV, this behavior is well reproduced by a Damped Random Walk (DRW) process (see Kelly et al., 2009), interpreted as the natural outcome of turbulence and thermal fluctuations in a single accretion disk. In this framework, the temporal variability of a single AGN is statistically described by one DRW component. This recently motivates a new time-domain approach for identifying unresolved ultra-compact dual AGN (Bertassi et al., 2026): if two ultra-compact MBHs are present, each powered by its own accretion flow, the observed, spatially-unresolved, variability should arise from the superposition of two independent stochastic processes. Therefore, by comparing whether a light curve is better modeled by a single DRW (single AGN) or by two independent DRWs (dual AGN) within a Bayesian inference framework, it is possible to statistically unveil dual-AGN candidates even when the system is spatially unresolved. In the next few years, this approach will enable the identification of large samples of unresolved systems across cosmic time, using wide-field and/or high-angular-resolution time-domain surveys conducted by facilities such as Gaia (Evans et al., 2023), Euclid (Euclid Collaboration et al., 2025), Nancy Grace Roman Space Telescope (Perkins et al., 2024), and Large Synoptic Survey Telescope (LSST, Ivezić et al., 2019),

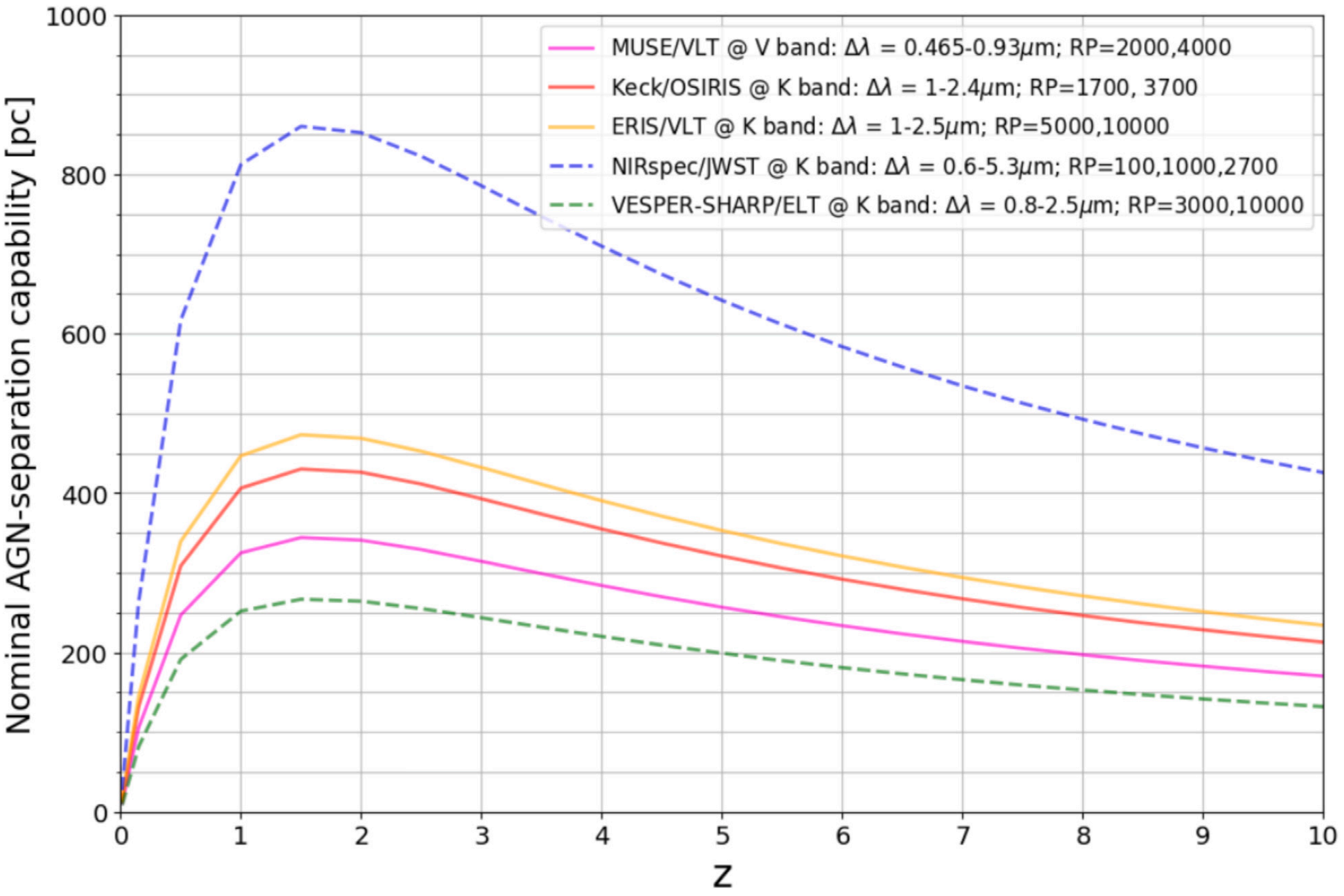

**Fig. 1.** Nominal AGN-separation capability (in pc) as a function of redshift for different observing facilities. The curves illustrate the minimum physical separation corresponding to the angular resolution element achievable in the diffraction-limited regime for each instrument. For ERIS and MUSE (VLT) and OSIRIS (Keck), the resolving capability is quantified using the AO-corrected FWHM of the PSF. For comparison, in the case of NIRSpec (JWST) and VESPER (ELT), whose diffraction-limited PSFs are undersampled, we adopt the pixel scale as an effective measure of the resolving capability. We also report the spectral range covered by the instruments, $\delta\lambda$, and their resolving power, $RP$.

flagging candidate dual AGN with projected separations potentially down to a few parsecs.

However, identifying candidates is only the first step. Confirming the presence of two AGN is significantly more challenging, as it requires resolving the pair at very small projected separations (see e.g., De Rosa et al., 2019; Severgnini et al., 2021; Zheng et al., 2024; Trindade Falcão et al., 2024; Lamperti et al., 2026). With current facilities, integral-field spectroscopy can separate and study two nuclei at hundreds of pc of separation only at z≲0.5, where angular resolution (pixel scales~0.1-0.2 arcsec) and sensitivity are still adequate. At higher redshift, these sources remain unresolved, and their spectroscopic signatures are impossible to disentangle, preventing secure confirmation and detailed characterization. The situation for double-DRW selected AGN is even more limiting from an observational point of view: at present, it is not possible to spatially resolve and characterize two accreting nuclei at separations of a few parsecs, even at z~0.1.

As a consequence, despite the growing number of sub-kpc dual AGN candidates identified so far, the confirmed fraction in the local Universe remains roughly an order of magnitude below theoretical predictions (e.g., Foreman et al., 2009; Solanes et al., 2019) and no sub-kpc dual AGN has yet been found at high redshift (Chen et al., 2022; Gross et al., 2025).

## 4. The role of SHARP

### 4.1. Expected performance

SHARP is a multi-mode spectrograph consisting of two units: the multi-Integral Field Unit (m-IFU), VESPER, and the Multi-Object Spectrograph, NEXUS. While VESPER provides two-dimensional spatially resolved spectroscopy (twelve probes called Integral Field Selectors, each with a field of view of $1.7'' \times 1.5''$), NEXUS can deliver spatially resolved spectroscopic information along each single slit, allowing both observing modes to be effectively exploited for the study of compact dual AGN. The high angular resolution achievable with SHARP (pixel size of ~31–35 mas), when coupled with a multi-conjugate adaptive optics system such as MORFEO@ELT, provides direct access to the ultra-compact dual AGN regime. This capability, combined with the high sensitivity expected for SHARP@ELT (Saracco et al., 2024), will enable the physical and kinematic characterization of both nuclei with projected relative separations below a few hundred parsecs at all redshifts (at least for not too-obscured systems), and potentially down to a few pc in the local Universe.

Fig. 1 presents the nominal AGN-separation capability as a function of redshift for several observing facilities, including those on JWST, 8–10 m class telescopes, and the ELT. The curves represent the angular resolution element under ideal observing conditions for each facility, i.e. the best performance achievable in the diffraction-limited case (see also Davies et al. 2023, Jorgenson and Wolfe 2014 for actual performances of ERIS and OSIRIS, respectively). This comparison highlights the improvement in spatial resolving power from current large ground- and space-based facilities to next-generation extremely large telescopes, and quantifies their capability to resolve compact sources at increasing redshift.

The high angular capabilities of SHARP in resolving pc-scale separation sources in the local Universe are evident from the simulation shown in Fig. 2. The combination of high spatial resolution and IFU spectroscopy will overcome the difficulties of distinguishing nuclear emission, in particular in the case of extinction, from the dominant emission of the host galaxy. Diagnostics based on spatially-resolved emission-line ratios (see Section 4.2 for a list of expected emission lines) will enhance the ability to distinguish AGN emission from star-formation processes.

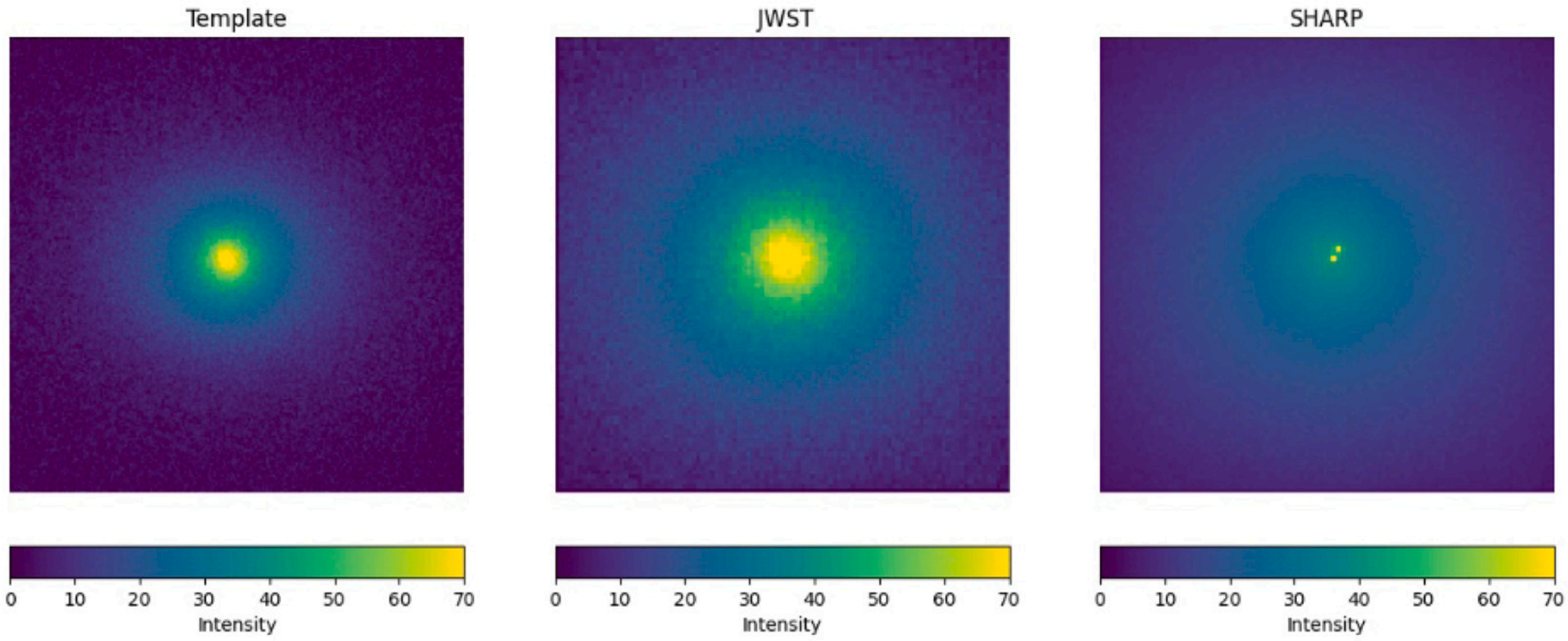

**Fig. 2.** Simulated observations of a template host galaxy with an embedded dual AGN system. Left panel: original galaxy template from LUCI (Large Binocular Telescope) observation in K band. Middle panel: simulated JWST observation, where the two AGN are unresolved and appear as a single central source. Right panel: simulated SHARP observation, where the two AGN are clearly resolved. In this example, the AGN separation is set to 2 pixels, corresponding to a physical separation of ~10 pc at z~0.01, illustrating the resolving power gain achieved with SHARP compared to JWST.

In this regard, in Fig. 2 two low-luminosity AGN (bolometric luminosity of a few $\times 10^{39}$ erg s$^{-1}$) are embedded in the bright host-galaxy emission (K = 13 mag, corresponding to a $\nu L_\nu(K) \sim 10^{45}$ erg s$^{-1}$ at z = 0.1). While JWST is not able to resolve the two nuclei, SHARP will detect them on top of the extended emission. As discussed above, resolving pair of AGN at such parsec-scale separations is currently extremely challenging: to date, only one confirmed system hosting two AGN at ~7 pc separation is known (Rodriguez et al., 2006).

To quantify the range of systems that can be observed, we estimate the expected K-band apparent magnitudes as a function of the AGN bolometric luminosity and redshift. We derived the near-IR bolometric correction using the relation of Spinoglio et al. (2024). The corresponding AGN apparent magnitudes for $L_{bol} = 10^{43}$, $10^{44}$ and $10^{45}$ erg s$^{-1}$ are reported in Table 1 for representative redshifts $z$ = 1, 2, 4, and 8. Using the SHARP@ELT exposure time calculator (ETC, v. 0.6)[3] and adopting a spectral resolution of $R$ = 3000, we estimated that for a typical AGN with $L_{bol} \sim 10^{45}$ erg s$^{-1}$, corresponding to $K_{AB} \leq 24$ even at $z \sim 8$ (see Table 1), a $S/N \geq 10$ per spectral resolution element can be achieved for the continuum emission in less than one hour of integration time. This sensitivity will allow even the detection of faint emission lines (EW ~ 5 Å). For weaker sources ($K_{AB} \sim 26.5$, corresponding to AGN with $L_{bol} \sim 10^{43-44}$ erg s$^{-1}$ at intermediate/high redshift; see Table 1), a $S/N \sim 3$ of the continuum emission can still be reached within ~7 hours, enabling the detection of stronger emission-line features. This implies that the high sensitivity and throughput of VESPER (and NEXUS) will allow building statistically large and reasonably complete samples of compact dual AGN, including obscured and low-contrast pairs, paving the way toward population studies and demographic trends across cosmic time.

Targets for SHARP will be drawn from the most complete and unbiased catalogues of sub-kpc dual AGN that will be available in the next years. These include the rapidly expanding samples of double-peaked AGN identified in spectroscopic surveys (see Section 3.2). In particular, at low redshift (z<0.15), double-peaked emission line AGN selected from SDSS (Sloan Digital Sky Survey; York et al., 2000) and lacking evidence for double stellar cores in existing sub-arcsecond optical or NIR imaging represent prime candidates; the absence of resolved nuclei could actually imply extremely small separations.

[3] https://sharp.lambrate.inaf.it/. At the time of writing, the ETC is still undergoing beta testing.

**Table 1**
$K$-band apparent AB-magnitude for AGN of different bolometric luminosities as a function of redshift. No $K$-correction or extinction has been applied.

| $L_{bol}$ | K-band (AB mag) | | | |
|---|---|---|---|---|
| (erg s$^{-1}$) | ($z$ = 1) | ($z$ = 2) | ($z$ = 4) | ($z$ = 8) |
| $10^{43}$ | 23.3 | 25.2 | 26.9 | 28.7 |
| $10^{44}$ | 20.9 | 22.8 | 24.6 | 26.4 |
| $10^{45}$ | 18.6 | 20.4 | 22.3 | 24 |

At higher redshift, double-peaked emission-line AGN identified in SDSS and GNIRS-DQS (Gemini Near Infrared Spectrograph - Distant Quasar Survey; Matthews et al., 2021) and associated with a single, morphologically unperturbed galaxy (using SDSS or HST imaging) are expected to host sub-kiloparsec MBH pairs. In parallel, double-DRW variability diagnostic will provide an independent and largely unbiased route to identify ultra-compact dual AGN beyond the limits of angular resolution. The synergy between double-peaked AGN and double-DRW selections will thus yield the most complete catalogue of compact dual AGN candidates available for SHARP follow-up.

### 4.2. SHARP science on ultra-compact-dual AGN

SHARP will enable the direct measurement of a set of physical quantities fundamental for the confirmation and characterization of ultra-compact dual AGN, as detailed below:

• *Detection of double nuclei and estimate of their physical separation:* the physical separations will be derived by combining spatial projected separation and spectral emission line information.

• *Emission-line profiles and FWHM diagnostics:* measured FWHM values will classify individual nuclei as broad emission-line (FWHM>1500 km/s, i.e. unobscured AGN) or narrow emission-line sources (AGN or HII regions).

•*Emission-line fluxes and line-ratio diagnostics:* flux measurements of rest-frame optical/NIR lines accessible with SHARP (e.g. H$\alpha$, H$\beta$, [OIII]$\lambda$5007, [NII]$\lambda$6584, [SII]$\lambda$6716, 6731, [SIII]$\lambda$9069,9532, Pa$\beta$, Pa$\alpha$, [FeII], Br$\gamma$, $H_2$, see e.g. Calabrò et al. 2023) will enable AGN–SF separation, gas excitation and ionization mapping, and metallicity and extinction estimates in each nucleus independently. SHARP diagnostics will be highly complementary to mid-IR observations with JWST/MIRI, which probe dust-obscured AGN activity, PAH emission, and warm molecular gas through ionic and molecular features (e.g., García-Bernete et al., 2024). In addition, synergy with high-resolution ALMA CO, HCN, and $HCO^+$ maps (tracing gas at different densities) will provide a multi-phase view of the gas content and kinematics in merging systems, linking the ionized and warm molecular gas observed by SHARP to the cold molecular reservoir (e.g., Davies et al., 2012; Cicone et al., 2018; Butterworth et al., 2025; Hagedorn et al., 2026).

• *BH mass and luminosity estimates:* continuum and broad emission-line fluxes will yield black hole mass and Eddington ratio estimates through scaling relations (e.g., Vestergaard and Peterson, 2006; McConnell and Ma, 2013), providing accretion rate and evolutionary stage information for each nucleus.

• *Morphology and kinematics of nuclear gas:* the datacube will enable 3D modeling of ionized and molecular gas (e.g. BAROLO 3D, Di Teodoro and Fraternali 2015; GalPaK 3D, Bouché et al. 2015; MOKA$^{3D}$, Marconcini et al. 2023) to recover rotation curves, inflows/outflows signatures, and velocity dispersion maps at parsec resolution.

• *Star formation and shocked gas diagnostics:* line ratios involving [OIII], H recombination lines, [FeII], $H_2$ and Br$\gamma$/Pa$\beta$ will probe star formation, shock excitation and feedback with unprecedented spatial resolution, tracing how MBH-driven winds impact the nuclear ISM.

• *Fraction of confirmed dual AGN:* by combining spatial resolution and spectral diagnostics, SHARP will convert statistical double-peaked emission-line and double-DRW sources into confirmed ultra-compact dual AGN, allowing demographic studies and the first robust measurement of the dual fraction at scales ranging from hundreds of pc to pc.

## 5. Resolving the key open questions with SHARP

While different photometric and spectroscopic techniques already enable the construction of statistical samples of ultra-compact dual-AGN candidates, the confirmation and the physical characterization of these systems require resolving the two accretion flows and their emission-line regions, disentangling spectra, mapping velocity fields and measuring extinction, metallicity and feedback. Such observations are effectively inaccessible today for ultra-compact systems up to cosmological distances.

SHARP@ELT is expected to fundamentally advance this field, finally enabling the detection and spectroscopic characterization of ultra-compact dual AGN across cosmic time, turning candidates into physically confirmed systems. This will finally allow us to address some of the still open key questions in MBH-galaxy co-evolution:

**- Are galaxy mergers the main drivers of MBH assembly and growth?**

By building the first large, unbiased sample of confirmed ultra-compact dual AGN, we will be able to quantify the incidence of dual accretion in merging galaxies, its evolution across redshift, and its role within MBH growth pathways.

**- What physical processes regulate the AGN–galaxy interplay in the final stages of mergers?**

SHARP will allow us to characterize ionized gas, highly-ionized coronal gas and dust in each nucleus independently. This will enable robust measurements of winds/outflows, feeding and feedback, and provide direct constraints on the physics of accretion and ejection in interacting MBH systems.

**- What are the electromagnetic precursors of low–frequency GW sources?**

By identifying and characterizing compact dual AGN down to pc scales, SHARP@ELT will reveal the immediate progenitors of binary MBHs that will be detectable by LGWA, LISA and PTA, allowing us to connect the electromagnetic properties of compact AGN pairs with the properties of their GW-emitting descendants. In particular, SHARP will provide key electromagnetic observables including black hole masses (via virial estimators when broad emission lines are present; see, e.g. Blandford and McKee 1982, Peterson 1993, Kaspi et al. 2000, Peterson et al. 2004, Shen et al. 2008), Eddington ratios, gas kinematics, and outflow energetics in both nuclei. These quantities define the physical conditions of MBH pairs prior to binary formation and regulate their interaction with the surrounding stellar and gaseous environment (e.g. via dynamical friction, stellar hardening, and gas torques), which in turn governs the efficiency of orbital decay (e.g. Peters, 1964; Begelman et al., 1980; Dotti et al., 2012; Mayer, 2013; Sesana, 2013).

Moreover, the number density and separation distribution of dual AGN in the ∼1–1000 pc regime will provide a direct empirical constraint on the timescale for MBH binary formation (i.e. the “delay time” between dual AGN and coalescing binaries). This timescale is a key parameter in current models of MBH merger rates (e.g. Sesana, 2013; Kelley et al., 2017; Bonetti et al., 2019; Singh et al., 2026; Davari et al., 2026), but remains highly uncertain. In this regard, SHARP observations will translate into improved estimates of MBH merger rates and allow us to place forthcoming low-frequency GW detections into a significantly more robust astrophysical context.

## CRediT authorship contribution statement

**P. Severgnini:** Writing – review & editing, Writing – original draft, Methodology, Conceptualization. **C. Vignali:** Writing – review & editing, Writing – original draft, Methodology, Conceptualization. **A. De Rosa:** Writing – review & editing. **E. Portaluri:** Writing – review & editing. **F. Rigamonti:** Writing – review & editing. **L. Battistini:** Writing – review & editing. **L. Bertassi:** Writing – review & editing. **E. Bertola:** Writing – review & editing. **S. Bianchi:** Writing – review & editing. **E. Bortolas:** Writing – review & editing. **C. Cicone:** Writing – review & editing. **Q. D'Amato:** Writing – review & editing. **R. Della Ceca:** Writing – review & editing. **I. Delvecchio:** Writing – review & editing. **M. Dotti:** Writing – review & editing. **J. Harms:** Writing – review & editing. **I. Lamperti:** Writing – review & editing. **F. Mannucci:** Writing – review & editing. **M. Parvatikar:** Writing – review & editing. **B. Sala:** Writing – review & editing. **M. Scialpi:** Writing – review & editing. **R. Serafinelli:** Writing – review & editing. **J. Singh:** Writing – review & editing. **M.V. Zanchettin:** Writing – review & editing.

## Declaration of competing interest

The authors declare that they have no known competing financial interests or personal relationships that could have appeared to influence the work reported in this paper.

## Acknowledgments

We thank the anonymous referee for the comments and suggestions. We acknowledge a financial contribution from the Bando Ricerca Fondamentale INAF 2022 Large Grant, 'Dual and binary supermassive black holes in the multi-messenger era: from galaxy mergers to gravitational waves' and the Bando Ricerca Fondamentale INAF 2024 Large Grant, 'The Quest for dual and binary massive black holes in the gravitational wave era'. We acknowledge financial support from the Italian Space Agency under Grant No. 2025-29-HH.0 and No. 2024-36-HH.1-2025. EB acknowledges INAF, Bando Ricerca Fondamentale INAF 2024 GO grant “A JWST/MIRI MIRACLE: Mid-IR Activity of Circumnuclear Line Emission” and Ricerca Fondamentale INAF 2024 MiniGrant RSN1 1.05.24.07.01. EP acknowledges the Bando di Ricerca Fondamentale INAF 2022 MiniGrant RSN1. RS acknowledges funding from the CAS-ANID grant number CAS220016.

## Data availability

No data was used for the research described in the article.

## References

Ajith, P., Seoane, P.A., Arca Sedda, M., et al., 2025. J. Cosmol. Astropart. Phys. 2025 (1), 108.
Amaro-Seoane, P., Andrews, J., Arca Sedda, M., et al., 2023. Living Rev. Relativ. 26 (1), 2.
Bacon, R., Accardo, M., Adjali, L., et al., 2010. In: McLean, I.S., Ramsay, S.K., Takami, H. (Eds.), Ground-Based and Airborne Instrumentation for Astronomy III. In: Society of Photo-Optical Instrumentation Engineers (SPIE) Conference Series, vol. 7735, 773508.
Bandara, K., Crampton, D., Simard, L., 2009. Astrophys. J. 704 (2), 1135–1145.
Barrows, R.S., Comerford, J.M., Stern, D., Assef, R.J., 2023. Astrophys. J. 951 (2), 92.
Battistini, L., De Rosa, A., Severgnini, P., et al., 2026. Astron. Astrophys. 710, A232.
Begelman, M.C., Blandford, R.D., Rees, M.J., 1980. Nature 287 (5780), 307–309.
Bertassi, L., Charisi, M., Rigamonti, F., Covino, S., Dotti, M., 2026. Astron. Astrophys. 709, A246.
Blandford, R.D., McKee, C.F., 1982. Astrophys. J. 255, 419–439.
Blecha, L., Loeb, A., Narayan, R., 2013. Mon. Not. R. Astron. Soc. 429 (3), 2594–2616.
Bonetti, M., Sesana, A., Barausse, E., Haardt, F., 2018. Mon. Not. R. Astron. Soc. 477 (2), 2599–2612.
Bonetti, M., Sesana, A., Haardt, F., Barausse, E., Colpi, M., 2019. Mon. Not. R. Astron. Soc. 486 (3), 4044–4060.
Bouché, N., Carfantan, H., Schroetter, I., Michel-Dansac, L., Contini, T., 2015. Astron. J. 150 (3), 92.
Buchner, J., Treister, E., Bauer, F.E., Sartori, L.F., Schawinski, K., 2019. Astrophys. J. 874 (2), 117.
Butterworth, J., Viti, S., Wang, Y., 2025. Astron. Astrophys. 697, A18.
Calabrò, A., Pentericci, L., Feltre, A., et al., 2023. Astron. Astrophys. 679, A80.
Carlsen, J., Cicone, C., Hagedorn, B., et al., 2026. Astron. Astrophys. 709, A134.
Chen, Y.-C., Gross, A.C., Liu, X., et al., 2025. Astrophys. J. 988 (1), 126.
Chen, Y.-C., Hwang, H.-C., Shen, Y., et al., 2022. Astrophys. J. 925 (2), 162.
Chen, Y.-C., Liu, X., Foord, A., et al., 2023. Nature 616 (7955), 45–49.
Cicone, C., Severgnini, P., Papadopoulos, P.P., et al., 2018. Astrophys. J. 863 (2), 143.
Ciliegi, P., Agapito, G., Aliverti, M., et al., 2024. In: Jackson, K.J., Schmidt, D., Vernet, E. (Eds.), Adaptive Optics Systems IX. In: Society of Photo-Optical Instrumentation Engineers (SPIE) Conference Series, vol. 13097, 1309722.
Cisternas, M., Jahnke, K., Inskip, K.J., et al., 2011. Astrophys. J. 726 (2), 57.
Ciurlo, A., Mannucci, F., Yeh, S., et al., 2023. Astron. Astrophys. 671, L4.
Cole, S., Lacey, C.G., Baugh, C.M., Frenk, C.S., 2000. Mon. Not. R. Astron. Soc. 319 (1), 168–204.
Colpi, M., 2014. Space Sci. Rev. 183 (1–4), 189–221.
Colpi, M., Dotti, M., 2011. Adv. Sci. Lett. 4 (2), 181–203.
Comerford, J.M., Pooley, D., Barrows, R.S., et al., 2015. Astrophys. J. 806 (2), 219.
Cuillandre, J.-C., Bolzonella, M., Boselli, A., et al., 2025. Astron. Astrophys. 697, A11.
D'Amato, Q., Mannucci, F., Sonnenfeld, A., et al., 2026. Nat. Astron. 10, 901–912.
Davari, N., Valiante, R., Trinca, A., Schneider, R., Caleno, R., Colpi, M., Arca Sedda, M., Bonetti, M., Lupi, A., Decarli, R., Sesana, A., 2026. Gravitational Waves from the Cosmic Dawn: Tracing Cosmic Black Hole Binaries with ET, LGWA and LISA. arXiv e-prints, arXiv:2604.18173, http://dx.doi.org/10.48550/arXiv.2604.18173.
Davies, R., Absil, O., Agapito, G., et al., 2023. Astron. Astrophys. 674, A207.
Davies, R., Mark, D., Sternberg, A., 2012. Astron. Astrophys. 537, A133.
De Rosa, A., Vignali, C., Bogdanović, T., et al., 2019. New A Rev. 86, 101525.
De Rosa, A., Vignali, C., Husemann, B., et al., 2018. Mon. Not. R. Astron. Soc. 480 (2), 1639–1655.
De Rosa, A., Vignali, C., Severgnini, P., et al., 2023. Mon. Not. R. Astron. Soc. 519 (4), 5149–5160.
Deane, R.P., Paragi, Z., Jarvis, M.J., et al., 2014. Nature 511 (7507), 57–60.
Di Matteo, T., Springel, V., Hernquist, L., 2005. Nature 433 (7026), 604–607.
Di Teodoro, E.M., Fraternali, F., 2015. Mon. Not. R. Astron. Soc. 451 (3), 3021–3033.
Dotti, M., Sesana, A., Decarli, R., 2012. Adv. Astron. 2012, 940568.
Ellison, S.L., Patton, D.R., Mendel, J.T., Scudder, J.M., 2011. Mon. Not. R. Astron. Soc. 418 (3), 2043–2053.
Ellison, S.L., Patton, D.R., Simard, L., McConnachie, A.W., 2008. Astron. J. 135 (5), 1877–1899.
EPTA Collaboration, InPTA Collaboration, Antoniadis, J., et al., 2023. Astron. Astrophys. 678, A50.
Eracleous, M., Boroson, T.A., Halpern, J.P., Liu, J., 2012. Astrophys. J. Suppl. 201, 23.
Eracleous, M., Halpern, J.P., M. Gilbert, A., Newman, J.A., Filippenko, A.V., 1997. Astrophys. J. 490 (1), 216–226.
Euclid Collaboration, Mellier, Y., Abdurro'uf, et al., 2025. Astron. Astrophys. 697, A1.
Evans, D.W., Eyer, L., Busso, G., et al., 2023. Astron. Astrophys. 674, A4.
Fabricius, M., Saglia, R., Balzer, F., et al., 2025. Euclid: Quick Data Release (Q1) – Secondary nuclei in early-type galaxies. arXiv e-Prints, arXiv:2511.02988.
Ferrarese, L., Ford, H., 2005. Space Sci. Rev. 116 (3–4), 523–624.
Ferrarese, L., Merritt, D., 2000. Astrophys. J. Lett. 539 (1), L9–L12.
Foreman, G., Volonteri, M., Dotti, M., 2009. Astrophys. J. 693 (2), 1554–1562.
Fu, H., Myers, A.D., Djorgovski, S.G., et al., 2015. Astrophys. J. 799 (1), 72.
Fu, H., Yan, L., Myers, A.D., et al., 2012. Astrophys. J. 745 (1), 67.
Gabányi, K.É., Frey, S., Paragi, Z., An, T., Komossa, S., 2017. In: Gomboc, A. (Ed.), New Frontiers in Black Hole Astrophysics. In: IAU Symposium, 324, pp. 223–226.
García-Bernete, I., Rigopoulou, D., Donnan, F.R., et al., 2024. Astron. Astrophys. 691, A162.
Gaskell, C.M., Harrington, P.Z., 2018. Mon. Not. R. Astron. Soc. 478 (2), 1660–1669.
Ge, J.-Q., Hu, C., Wang, J.-M., Bai, J.-M., Zhang, S., 2012. Astrophys. J. Suppl. 201 (2), 31.
Gebhardt, K., Bender, R., Bower, G., et al., 2000. Astrophys. J. Lett. 539 (1), L13–L16.
Glikman, E., Langgin, R., Johnstone, M.A., et al., 2023. Astrophys. J. Lett. 951 (1), L18.
Gravity Collaboration, Abuter, R., Accardo, M., et al., 2017. Astron. Astrophys. 602, A94.
Gravity+ Collaboration, Abuter, R., Allouche, F., et al., 2026. Astron. Astrophys. 707, A115.
Gross, A.C., Chen, Y.-C., Oguri, M., et al., 2025. Astrophys. J. 989 (1), 112.
Hagedorn, B., Cicone, C., Sarzi, M., Severgnini, P., Vignali, C., 2026. Astron. Astrophys. 707, A77.
Harms, J., Ambrosino, F., Angelini, L., et al., 2021. Astrophys. J. 910 (1), 1.
Heckman, T.M., Miley, G.K., Green, R.F., 1984. Astrophys. J. 281, 525–534.
Hopkins, P.F., Hernquist, L., Cox, T.J., Kereš, D., 2008. Astrophys. J. Suppl. 175 (2), 356–389.
Hopkins, P.F., Hernquist, L., Cox, T.J., et al., 2006. Astrophys. J. Suppl. 163 (1), 1–49.
Hou, M., Li, Z., Liu, X., 2020. Astrophys. J. 900 (1), 79.
Hou, M., Liu, X., Guo, H., et al., 2019. Astrophys. J. 882 (1), 41.
Hwang, H.-C., Shen, Y., Zakamska, N., Liu, X., 2020. Astrophys. J. 888 (2), 73.
Ivezić, v., Kahn, S.M., Tyson, J.A., et al., 2019. Astrophys. J. 873 (2), 111.
Jakobsen, P., Ferruit, P., Alves de Oliveira, C., et al., 2022. Astron. Astrophys. 661, A80.
Jin, G., Dai, Y.S., Pan, H.-A., et al., 2021. Astrophys. J. 923 (1), 6.
Jorgenson, R.A., Wolfe, A.M., 2014. Astrophys. J. 785 (1), 16.
Jovanović, P., Popović, L.v., Stalevski, M., Shapovalova, A.I., 2010. Astrophys. J. 718 (1), 168–176.
Ju, W., Greene, J.E., Rafikov, R.R., Bickerton, S.J., Badenes, C., 2013. Astrophys. J. 777, 44.
Kaspi, S., Smith, P.S., Netzer, H., et al., 2000. Astrophys. J. 533 (2), 631–649.
Kelley, L.Z., Blecha, L., Hernquist, L., 2017. Mon. Not. R. Astron. Soc. 471, 4508–4526.
Kelly, B.C., Bechtold, J., Siemiginowska, A., 2009. Astrophys. J. 698 (1), 895–910.
Kollatschny, W., Weilbacher, P.M., Ochmann, M.W., et al., 2020. Astron. Astrophys. 633, A79.
Kormendy, J., Ho, L.C., 2013. Annu. Rev. Astron. Astrophys. 51 (1), 511–653.
Kormendy, J., Richstone, D., 1995. Annu. Rev. Astron. Astrophys. 33, 581.
Koss, M., Mushotzky, R., Treister, E., et al., 2012. Astrophys. J. Lett. 746 (2), L22.
Lamperti, I., Mannucci, F., Bertola, E., et al., 2026. Astron. Astrophys. 710, A193.
Larkin, J., Barczys, M., Krabbe, A., et al., 2006. In: McLean, I.S., Iye, M. (Eds.), Ground-Based and Airborne Instrumentation for Astronomy. In: Society of Photo-Optical Instrumentation Engineers (SPIE) Conference Series, vol. 6269, p. 62691A.
Lemon, C., Anguita, T., Auger-Williams, M.W., et al., 2023. Mon. Not. R. Astron. Soc. 520 (3), 3305–3328.
Lemon, C.A., Auger, M.W., McMahon, R.G., 2019. Mon. Not. R. Astron. Soc. 483 (3), 4242–4258.
Lemon, C.A., Auger, M.W., McMahon, R.G., Ostrovski, F., 2018. Mon. Not. R. Astron. Soc. 479 (4), 5060–5074.
Liu, X., Civano, F., Shen, Y., et al., 2013. Astrophys. J. 762 (2), 110.
Liu, X., Shen, Y., Strauss, M.A., Greene, J.E., 2010. Astrophys. J. 708 (1), 427–434.
Liu, X., Shen, Y., Strauss, M.A., Hao, L., 2011. Astrophys. J. 737 (2), 101.
Magorrian, J., Tremaine, S., Richstone, D., et al., 1998. Astron. J. 115 (6), 2285–2305.
Mahmoodzadeh, H., Saracco, P., Conconi, P., et al., 2025. J. Astron. Telesc. Instruments, Syst. 11, 035002.
Mannucci, F., Pancino, E., Belfiore, F., et al., 2022. Nat. Astron. 6, 1185–1192.
Mannucci, F., Scialpi, M., Ciurlo, A., et al., 2023. Astron. Astrophys. 680, A53.
Marconcini, C., Marconi, A., Cresci, G., et al., 2023. Astron. Astrophys. 677, A58.
Matthews, B.M., Shemmer, O., Dix, C., et al., 2021. Astrophys. J. Suppl. 252 (2), 15.
Mayer, L., 2013. Classical Quantum Gravity 30 (24), 244008.
McConnell, N.J., Ma, C.-P., 2013. Astrophys. J. 764 (2), 184.
McGurk, R.C., Max, C.E., Medling, A.M., Shields, G.A., Comerford, J.M., 2015. Astrophys. J. 811 (1), 14.
Mezcua, M., Lobanov, A.P., Mediavilla, E., Karouzos, M., 2014. Astrophys. J. 784 (1), 16.
Navarro, J.F., Frenk, C.S., White, S.D.M., 1996. Astrophys. J. 462, 563.
Nevin, R., Comerford, J., Müller-Sánchez, F., Barrows, R., Cooper, M., 2016. Astrophys. J. 832 (1), 67.
Paolillo, M., Papadakis, I., 2025. Nuovo Cimento Riv. Ser. 48 (8), 537–621.
Perkins, J.S., Wollack, E.J., Content, D.A., et al., 2024. In: Coyle, L.E., Matsuura, S., Perrin, M.D. (Eds.), Space Telescopes and Instrumentation 2024: Optical, Infrared, and Millimeter Wave. In: Society of Photo-Optical Instrumentation Engineers (SPIE) Conference Series, 13092, p. 130920R.
Perna, M., Arribas, S., Lamperti, I., et al., 2025. Astron. Astrophys. 696, A59.
Perna, M., Arribas, S., Marshall, M., et al., 2023. Astron. Astrophys. 679, A89.
Peters, P.C., 1964. Phys. Rev. 136 (4B), 1224–1232.
Peterson, B.M., 1993. Publ. Astron. Soc. Pac. 105, 247.
Peterson, B.M., Ferrarese, L., Gilbert, K.M., et al., 2004. Astrophys. J. 613 (2), 682–699.
Pfeifle, R.W., Satyapal, S., Manzano-King, C., et al., 2019a. Astrophys. J. 883 (2), 167.
Pfeifle, R.W., Satyapal, S., Secrest, N.J., et al., 2019b. Astrophys. J. 875 (2), 117.

Pfeifle, R.W., Weaver, K.A., Secrest, N.J., Rothberg, B., Patton, D.R., 2025. Astrophys. J. Suppl. 281 (1), 25.
Rigamonti, F., Severgnini, P., Sottocorno, E., et al., 2025. Astron. Astrophys. 693, A117.
Rodriguez, C., Taylor, G.B., Zavala, R.T., et al., 2006. Astrophys. J. 646 (1), 49–60.
Rosario, D.J., Shields, G.A., Taylor, G.B., Salviander, S., Smith, K.L., 2010. Astrophys. J. 716 (1), 131–143.
Sanders, D.B., Soifer, B.T., Elias, J.H., et al., 1988. Astrophys. J. 325, 74.
Saracco, P., Conconi, P., Arcidiacono, C., et al., 2024. In: Bryant, J.J., Motohara, K., Vernet, J.R.D. (Eds.), Ground-Based and Airborne Instrumentation for Astronomy X. In: Society of Photo-Optical Instrumentation Engineers (SPIE) Conference Series, vol. 13096, p. 130965I.
Satyapal, S., Secrest, N.J., Ricci, C., et al., 2017. Astrophys. J. 848 (2), 126.
Savorgnan, G.A.D., Graham, A.W., Marconi, A., Sani, E., 2016. Astrophys. J. 817 (1), 21.
Schawinski, K., Koss, M., Berney, S., Sartori, L.F., 2015. Mon. Not. R. Astron. Soc. 451 (3), 2517–2523.
Schwartzman, E., Fudolig, P., Clarke, T.E., et al., 2025. Astrophys. J. 987 (2), 200.
Scialpi, M., Mannucci, F., D'Amato, Q., et al., 2026. Astron. Astrophys. 709, A236.
Scialpi, M., Mannucci, F., Marconcini, C., et al., 2024. Astron. Astrophys. 690, A57.
Scudder, J.M., Ellison, S.L., Torrey, P., Patton, D.R., Mendel, J.T., 2012. Mon. Not. R. Astron. Soc. 426 (1), 549–565.
Sesana, A., 2013. Mon. Not. R. Astron. Soc. 433, L1–L5.
Severgnini, P., Braito, V., Cicone, C., et al., 2021. Astron. Astrophys. 646, A153.
Shen, Y., Greene, J.E., Strauss, M.A., Richards, G.T., Schneider, D.P., 2008. Astrophys. J. 680 (1), 169–190.
Shen, Y., Hwang, H.-C., Zakamska, N., Liu, X., 2019. Astrophys. J. Lett. 885 (1), L4.
Shen, Y., Liu, X., Greene, J.E., Strauss, M.A., 2011. Astrophys. J. 735 (1), 48.
Shen, Y., Liu, X., Loeb, A., Tremaine, S., 2013. Astrophys. J. 775, 49.
Singh, J., Severgnini, P., Cammelli, V., et al., 2026. Astron. Astrophys. 706, A196.
Smith, K.L., Shields, G.A., Bonning, E.W., et al., 2010. Astrophys. J. 716 (1), 866–877.
Solanes, J.M., Perea, J.D., Valentí-Rojas, G., et al., 2019. Astron. Astrophys. 624, A86.
Soltan, A., 1982. Mon. Not. R. Astron. Soc. 200, 115–122.
Sottocorno, E., Ogborn, M., Bertassi, L., et al., 2026. Astron. Astrophys. 708, A153.
Spingola, C., McKean, J.P., Massari, D., Koopmans, L.V.E., 2019. Astron. Astrophys. 630, A108.
Spinoglio, L., Fernández-Ontiveros, J.A., Malkan, M.A., 2024. Astrophys. J. 964 (2), 117.
Storchi-Bergmann, T., Schimoia, J.S., Peterson, B.M., et al., 2017. Astrophys. J. 835 (2), 236.
Treister, E., Schawinski, K., Urry, C.M., Simmons, B.D., 2012. Astrophys. J. Lett. 758 (2), L39.
Trindade Falcão, A., Turner, T.J., Kraemer, S.B., et al., 2024. Astrophys. J. 972 (2), 185.
Tsalmantza, P., Decarli, R., Dotti, M., Hogg, D.W., 2011. Astrophys. J. 738, 20.
Ulivi, L., Mannucci, F., Scialpi, M., et al., 2025. Euclid: A machine-learning search for dual and lensed AGN at sub-arcsec separations. arXiv e-Prints, arXiv:2508.19494.
Uppal, A., Ward, C., Gezari, S., et al., 2024. Astrophys. J. 975 (2), 286.
Van Wassenhove, S., Volonteri, M., Mayer, L., et al., 2012. Astrophys. J. Lett. 748 (1), L7.
Vestergaard, M., Peterson, B.M., 2006. Astrophys. J. 641 (2), 689–709.
Volonteri, M., Madau, P., 2008. Astrophys. J. Lett. 687 (2), L57.
Wang, J.-M., Chen, Y.-M., Hu, C., et al., 2009. Astrophys. J. Lett. 705 (1), L76–L80.
Wang, Q., Ding, X., Silverman, J., et al., 2026. Astrophys. J. 997 (2), 269.
Wang, L., Greene, J.E., Ju, W., et al., 2017. Astrophys. J. 834, 129.
Wang, H.-C., Wang, J.-X., Gu, M.-F., Liao, M., 2023. Mon. Not. R. Astron. Soc. 524 (1), L38–L44.
Ward, C., Koss, M.J., Eracleous, M., et al., 2025. Astrophys. J. 991 (1), 116.
Weigel, A.K., Schawinski, K., Treister, E., Trakhtenbrot, B., Sanders, D.B., 2018. Mon. Not. R. Astron. Soc. 476 (2), 2308–2317.
White, S.D.M., Frenk, C.S., 1991. Astrophys. J. 379, 52.
Whittle, M., Rosario, D.J., Silverman, J.D., Nelson, C.H., Wilson, A.S., 2005. Astron. J. 129 (1), 104–124.
Xu, W., Cui, L., Liu, X., et al., 2024. Astrophys. J. 969 (1), 36.
Xu, D., Komossa, S., 2009. Astrophys. J. Lett. 705 (1), L20–L24.
York, D.G., Adelman, J., Anderson, Jr., J.E., et al., 2000. Astron. J. 120 (3), 1579–1587.
Yoshida, M., Yagi, M., Ohyama, Y., et al., 2016. Astrophys. J. 820 (1), 48.
Zheng, Q., Zhang, X., Yuan, Q., Severgnini, P., Vignali, C., 2024. Mon. Not. R. Astron. Soc. 531 (1), L76–L81.